\documentclass[a4paper,12pt]{article}

\usepackage[latin1]{inputenc}   
\usepackage[francais]{babel}    
\usepackage{fancyhdr}       
\usepackage{graphicx}       
\usepackage{geometry}       
\usepackage{amsmath}        

\newcommand{\be}{\begin{equation}}
\newcommand{\ee}{\end{equation}}
\newcommand{\bc}{\begin{center}}
\newcommand{\ec}{\end{center}}
\newcommand{\glv}{\gamma_{LV}}

\usepackage{graphicx}
\begin{document}

\bc

{\bf \large Nucleation in hydrophobic cylindrical pores : a
lattice model}

\vspace{1cm}
         A. SAUGEY$^{a}$, L. BOCQUET$^{b}$, J.L. BARRAT$^{b}$

 \vspace{1cm}

 $^{a}$     \textit{Laboratoire de Tribologie et Dynamique des Syst\`emes, Ecole Centrale de Lyon and CNRS,
            36 Avenue Guy de Collongues, BP163, 69134 Ecully Cedex, France}\\
 $^{b}$     \textit{Laboratoire de Physique de la Mati\`ere Condens\'ee et Nanostructures,
            Universit\'e Claude Bernard Lyon I and CNRS, 6 rue Amp\`ere, 69622 Villeurbanne Cedex, France}

\vspace{1cm}

{\bf ABSTRACT}

\ec
 We consider the nucleation process associated
with capillary condensation of a vapor in a hydrophobic
cylindrical pore (capillary evaporation). The liquid-vapor
transition is described within the framework of a simple lattice
model. The phase properties are characterized both at the
mean-field level and using Monte-Carlo simulations. The nucleation
process for the liquid to vapor transition is then specifically
considered. Using umbrella sampling techniques, we show that
nucleation occurs through the condensation of an asymmetric vapor
bubble at the pore surface. Even for highly confined systems, good
agreement is found with macroscopic considerations based on
classical nucleation theory. The results are discussed in the
context of recent experimental work on the extrusion of water in
hydrophobic pores.

{\bf PACS:}


\section{Introduction}
Recently, the study of water confined in hydrophobic pores has
been the object of a growing interest, both from the fundamental
and the industrial point of view
\cite{maibaum,hummer,bolhuis,barrat,eros01,lefevre}. A specific
feature of such mesoporous materials is the strong adsorption of
the wetting phase occuring at a chemical potential (or pressure)
lower than the bulk saturation value. This behavior is usually
known as capillary condensation, and corresponds fundamentally to
the shifted liquid-gas phase transition induced by confinement
\cite{gelb}. In the case of hydrophobic pores, the wetting phase
is the vapor while the non wetting phase is the liquid. The
restricted geometry therefore favors nucleation of vapor bubbles
inside the pores. This is known as the hydrophobic effect in
chemistry, widely study by Chandler and co-workers
\cite{chandlernature,maibaum2,tenwolde,huang2}.

A characteristic of capillary condensation (both in hydrophobic
and hydrophilic porous matrices) is the existence of a large
hysteresis of adsorption. Two different origins have been pointed
out to explain this behavior. One is the presence  of large energy
barriers to nucleate the wetting phase (below its saturation
value) \cite{huang}. Another explanation, valid in particular for
disordered mesoporous matrices, is  the trapping of the system in
a complex free energy landscape \cite{kierlikjpcm,kierlikprl}.
There is in general an intricate coupling between these two
origins of hysteresis. However, limiting cases might be considered
experimentally. Highly disordered  mesoporous materials (such as
porous glasses or silica gels)  must be described
 using the approach of reference \cite{kierlikprl}. For "ideal" systems
 with regular pore shapes,   such as MCM-41, a more standard thermodynamic
 approach is appropriate \cite{lefevrecnt}.

In the following we focus on adsorption in such "ideal" materials.
The slit pore geometry, in which fluids are confined between a
pair of infinite plate, has first been considered in the
literature \cite{bolhuis,lum,lum2}. A macroscopic approach was
used by Restagno {\it et al.} to determine free energy barriers
\cite{restagno}, which proved to be very large.
 Talanquer {\it et al.} \cite{talanquer} used density functional theory (DFT)
 to tackle this problem and showed that the macroscopic
description yields results in quantitative agreement with DFT
provided the effect of line tension  is taken into account.
Nucleation path proposed along these approaches are in agreement
with those determined by molecular simulations using sampling
methods \cite{bolhuis} developed  by D. Chandler's group
\cite{bolhuis2,dellago}.

The experimentally relevant case of cylindrical pores has been
considered more recently by Kornev and Neimark \cite{kornev} and
Lefevre \emph{et al.} \cite{lefevrecnt} along the same lines. An
important difference with the slit case, however, is that  the
curvature of the pore may lead to a critical nucleus lacking axial
symmetry \cite{lefevrecnt}. The latter results have been compared
with experimental data on hydrophobic MCM-41 materials, showing
good agreement with the measured hysteresis and estimated
nucleation barriers.

In this paper, we investigate nucleation in a hydrophobic
cylindrical pore using a lattice functional density approach. Our
aim is twofold : (i) assess the pertinence of the macroscopic
theory to describe nucleation under  strong confinement; (ii)
consider the possibility of  nucleation paths involving asymmetric
nuclei, which we previously predicted on the basis of macroscopic
arguments to be energetically favorable \cite{lefevrecnt}. To
this end, we consider a very simple coarse grained  model, which
takes fluid-fluid and fluid-solid interactions into account at the
most simple level. Critical temperature and chemical potential at
coexistence as well as liquid-vapor interfacial tension and
contact angle value are computed using both mean field
calculations and Monte Carlo simulations. Umbrella sampling is
used to determine nucleation paths, critical nuclei and reduced
energy barriers.

\section{Model}
\label{sect:model}

\subsection{Microscopic Hamiltonian : mean-field and fluctuations}

Our model is defined  by the Hamiltonian proposed by Kierlik {\it
et al.} to describe a confined inhomogeneous fluid in contact with
an external reservoir of temperature T and chemical potential
$\mu$ \cite{kierlikpccp,kierlikjpcm,sarkisov}~:
\begin{eqnarray}
\lefteqn{
H({\rho_i})=k_B T \sum_i[\rho_i ln \rho_i + (\eta_i-\rho_i)ln(\eta_i-\rho_i)] -\mu \sum_i \rho_i }\nonumber\\
& & {}-w_{ff}\sum_{<ij>} \rho_i\rho_j -w_{mf}\sum_{<ij>}
[\rho_i(1-\eta_j) +(1-\eta_i)\rho_j] \label{eqn:Hamiltonian}
\end{eqnarray}
In this expression, $1-\eta_i$ are the (discrete) occupancy
variables for the matrix ($\eta_i=0,1$ for the matrix/fluid site)
and $\rho_i$ is the  local density of the fluid on a
three-dimensional (BCC) lattice ($\rho_i \in[0,1]$). The
fluid-fluid $w_{ff}$ and matrix-fluid $w_{mf}$ interactions only
act between nearest neighbors sites ($<ij>$) and the ratio
$w_{mf}/w_{ff}$ determines the wettability of the matrix. This
coarse-grained description leaves aside most of the microscopic
details of an actual solid-fluid system but allows extensive
simulations while retaining the main experimental and physical
ingredients of the system under consideration. In particular it is
sufficient to describe, at least qualitatively, the interplay
between volume and surface contributions of the free energy (here
in a curved, cylinder like, geometry). At this point, one may
notice that the above Hamiltonian is equivalent to the
coarse-grained version of the well known site-diluted Ising-Model
\cite{chandler}.

\begin{figure}
\begin{center}
\includegraphics[height=5cm]{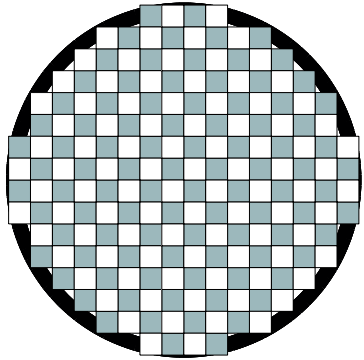}
\end{center}
\caption{Representation of the section
 of a cylindrical pore in the BCC lattice model. The sites shown
 belong to the fluid ($\eta_i=1$). The main axis of the cylinder
  is along a (100) direction the cubic lattice. The  radius of the cylinder
  is here 5.5 lattice spacings.}
\label{fig:reseau}
\end{figure}

The cylindrical pore is represented in figure \ref{fig:reseau},
together with the underlying BCC lattice. Periodic boundary
conditions are applied in the direction parallel to the axis of
the cylinder. The BCC lattice representation of a cylinder might
appear crude, but, as already mentioned above, this proves
sufficient to study the generic mechanism of nucleation in a
curved geometry.

We have investigated the phase properties of the above Hamiltonian
in equation \ref{eqn:Hamiltonian}, at two levels of description.
First at the mean field level, where the previous Hamiltonian is
identified as the free energy of the system; second, by performing
finite temperature Monte Carlo simulation of the Hamiltonian
(equation \ref{eqn:Hamiltonian}) in order to incorporate
fluctuation effects.

Before turning to the nucleation properties, we briefly
characterize the bulk and surface properties of the system.

\subsection{Bulk Phase properties}

\begin{figure}
\begin{center}
\includegraphics[height=8cm]{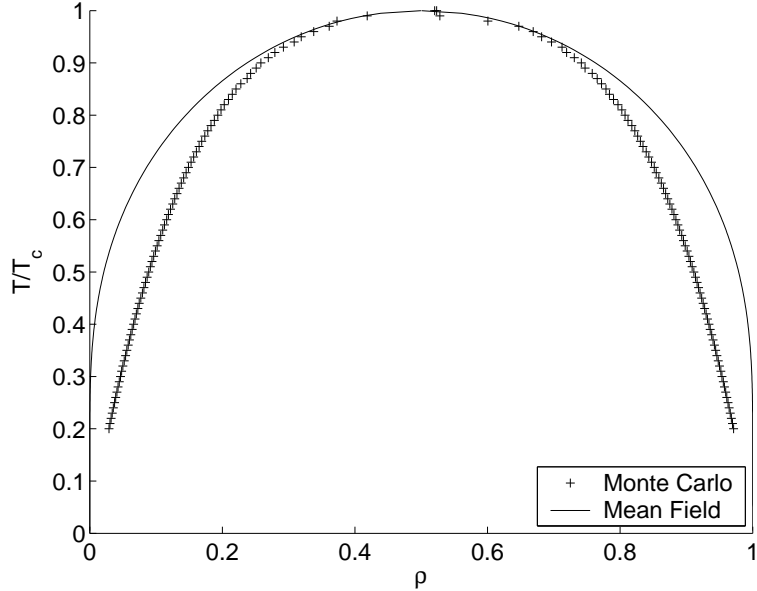}
\end{center}
\caption{Bulk phase diagram, obtained  from mean field (full
curve) and Monte-Carlo (dots) approaches.}
\label{fig:bulkphasdiag}
\end{figure}

In the bulk, the system undergoes a liquid-vapor phase transition.
At the mean field level, this is easily deduced from the equation
of state which is found to take the form~
 \begin{equation} \mu=k_B T \ln
\frac{\rho_{bulk}}{1-\rho_{bulk}}- z w_{ff} \rho_{bulk}
\end{equation} where $\rho_{bulk}$ is the uniform density of the
bulk fluid and $z=8$ is the lattice coordination number. The
mean-field bulk liquid-gas transition is then found to take place
at $\mu_{sat}/w_{ff}=-4.0$, with a critical point located at $k_B
T_c/w_{ff}=2.0$. The resulting mean-field bulk phase diagram is
plotted in figure \ref{fig:bulkphasdiag}.

Beyond mean field, fluctuations are incorporated by performing
Monte Carlo simulations based on the Hamiltonian in equation
\ref{eqn:Hamiltonian}. A few technical comments are in order here.
The standard Metropolis method is used \cite{chandler,frenkel}, in
the grand canonical ensemble (which amounts here to simply fixing
the chemical potential $\mu$ of the fluid). One Monte Carlo step
corresponds to one attempted trial move per fluid lattice site.
The  density change in the trial moves is $0.2$, corresponding to
a typical acceptance rate of $0.5$.

The bulk phase diagram is computed using periodic boundary
conditions, with a cubic simulation cell of  size $L$. The
liquid-vapor equilibrium is determined by equality of the grand
 potential in the two phases.
The latter is computed by thermodynamical integration of
\begin{equation}
\frac{\partial \Omega}{\partial \mu}=-<\rho>
\end{equation}
with boundary conditions
\begin{eqnarray}
\lim_{\mu \to -\infty}\Omega_v & = & 0  \\
\lim_{\mu \to +\infty}\Omega_l & = & -(\mu+4 w_{ff})
\end{eqnarray}
The simulated phase diagram is plotted in figure
\ref{fig:bulkphasdiag} (for a lattice size $L=15$, corresponding
to $2\times 15^3$ sites of the underlying BCC lattice).  The value
$\mu_{sat}/w_{ff}=-4.0$ gives a very good approximation of the
chemical potential on the critical line,  (independently of the
temperature $T$). The critical temperature is found to be $k_B
T_c/w_{ff}\simeq0.5$, quite smaller than the mean-field result
($k_B T_c/w_{ff}=2$). If fluctuations are expected to decrease the
critical temperature in spin-Ising model, such a large difference
is however surprising and seems to be due to the coarse graining
of the order parameter $\rho$.

\subsection{Surface properties}

In this section, we compute numerically the liquid vapor surface
tension and the contact angle of the triple line on the solid
surface. These ingredients will be needed to check the accuracy of
the  macroscopic calculation used to describe the nucleation
problem in  section \ref{nucleation}.

\subsubsection{Liquid-Vapor surface tension}
\label{sec:gammaLV}

Generally, the liquid vapor surface tension is computed by first constructing a liquid-
vapor interface and then computing the excess free energy (grand potential)
of this interface compared to the bulk coexistence free energy.
However in a lattice model, the surface free energy does depend on the particular
direction of the interface with respect to the underlying lattice axis.

\begin{figure}
\begin{center}
\includegraphics[height=8cm]{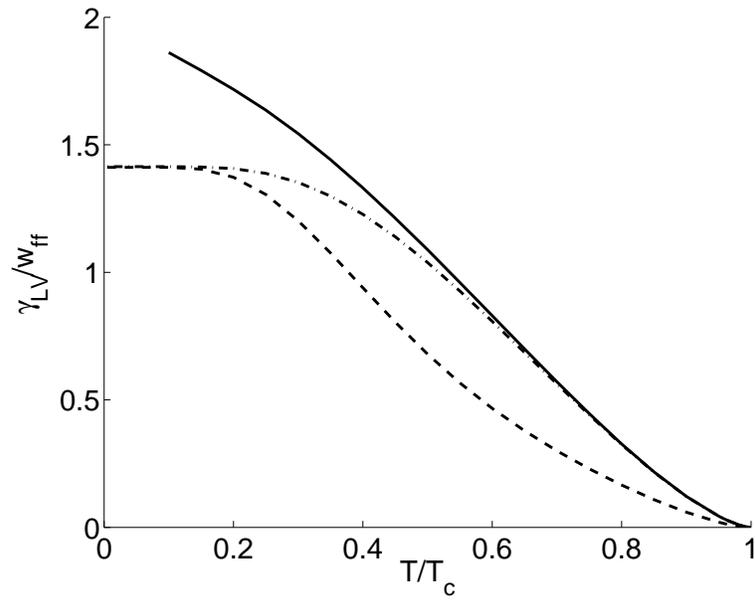}
\end{center}
\caption{Mean field estimate of Liquid-Vapor surface tension along
various directions, as a function of $T/T_c$. Solid line: surface
tension in the [100] direction, from  the excess grand potential.
Dash-dotted line: same estimate, now for the [110] interface.
 Dashed line:  estimate
from the mechanical route for the [110] interface.}
\label{fig:glv}
\end{figure}

This dependence is  emphasized below using the simple mean field
approach. In the presence of an interface,  minimization of the
(mean-field) Hamiltonian (\ref{eqn:Hamiltonian})  with respect to
the local fluid densities, yields a set of non linear coupled
equations \begin{equation}
\rho_i=\frac{\eta_i}{1+e^{-\beta\{\mu+\sum_{j/i} [w_{ff}\rho_j +
w_{mf}(1-\eta_j)]\}}}, \end{equation} where the sum is over the
nearest neighbors $j$ of site $i$. We have solved numerically
these equations by simple iterations
$\{\rho_i^{t+1}\}=f(\{\rho_i^t\})$ starting from an initial
distribution $\{\rho_i^0\}$ (with a convergence requirement
$\max_i |\rho_i^{t+1}-\rho_i^t|<10^{-10}$) with a sharp interface
in the desired orientation. Convergence is quite fast due to the
relative simple geometries considered here. Once the interfacial
profile is constructed, the mean field Grand Potential $\Omega$ is
simply computed by the value of the Hamiltonian computed at the
saddle point density. The surface tension is then defined as the
excess value, computed from the difference between this value and
the bulk coexistence grand potential. We plot on figure
\ref{fig:glv} the temperature dependence of $\gamma_{LV}$ for two
specific directions of the interface, along the [100] and [110]
planes. We also compare on this figure the value of the surface
tension obtained using a mechanical route, based on the dilation
of the sample volume, as described in the Appendix. Note the
difference between the thermodynamic and mechanical routes in the
present case, which can be ascribed here to the underlying
mean-field approximation \cite{rowlinsom}. 
Moreover another drawback of the mechanical estimate is that it is restricted
to the [110] interface, due to lattice effects.

This dependence on the lattice direction remains when fluctuations
are included, beyond the mean field approximation. Note however
that in contrast to the mean field case, the liquid-vapor surface
tension cannot be estimated from the excess grand
potential, which is not directly available in the simulation. We
have therefore first used the mechanical route, as described in the
appendix, to obtain the surface tension from the Monte-Carlo runs.
\begin{figure}
\begin{center}
\includegraphics[height=8cm]{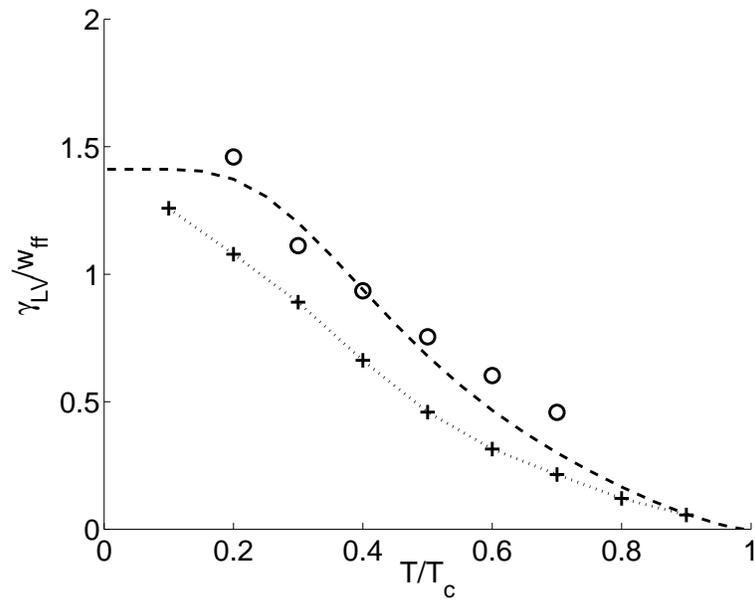}
\end{center}
\caption{Monte Carlo estimate of liquid-vapor surface tension as a
function of $T/T_c$.  Dotted line with crosses:   mechanical
estimate (implemented in the Monte Carlo sampling) in the [110]
direction.   Circles:  estimate from the free energy calculation
of a liquid bubble (see text for details). Dashed line: mean field
calculation for the [110] interface.} \label{fig:glvMC}
\end{figure}

A third route can  be proposed to estimate the liquid-vapor free
energy. Indeed the surface tension can be defined in terms of the
free energy necessary to create a bubble of vapor inside the
liquid, at coexistence. Moreover this estimate, which results
implicitly from an underlying average over the various direction
of the lattice grid, is particularly relevant when dealing with
the free energy of a nucleated bubble, which will be needed in the
next section.

The calculation of the free energy however requires a
thermodynamic integration. This is performed using the umbrella
sampling technique \cite{bolhuis2,frenkel}. At a given
temperature, the bulk (periodic) system  is placed at the
coexistence conditions, obtained from the previous section. An
order parameter $\Psi$ is then defined as the number of vapor
sites in the system ( with density less than $0.5$). We then
constrain the system to contain an average number of vapor sites
by biasing the bare Hamiltonian (\ref{eqn:Hamiltonian}). From the
technical point of view this is performed by adding a term in
$\frac{1}{2}\kappa(\Psi-\Psi_0)^2$ where $\Psi_0$ is the target
value for sampling. Typical values of $\kappa$ are $0.005$. The
Grand Potential curves is then deduced as a function of $\Psi$ by
computing  the state probability distribution $P(\Psi)$
\cite{frenkel}~:

\begin{equation} \label{umb} \Omega(\Psi)=-k_B T \ln
\left[P(\Psi)\exp\left({\frac{1}{2}\frac{\kappa(\Psi-\Psi_0)^2}{k_B
T}}\right)\right] \end{equation}
 A matching procedure for the free energy is required as
$\Psi_0$ is increased from $\Psi_0=0$ (liquid state) to $\Psi_0
\ne 0$ (vapor bubble). For a given order parameter, it can be
checked by inspecting the configurations that the vapor sites
organize into a vapor bubble with fixed radius $R$. It is then
possible to estimate the surface tension $\gamma_{LV}$ from the
definition $\Delta \Omega =4 \pi \gamma_{LV} R^2$.
\begin{figure}
\begin{center}
\includegraphics[height=8cm]{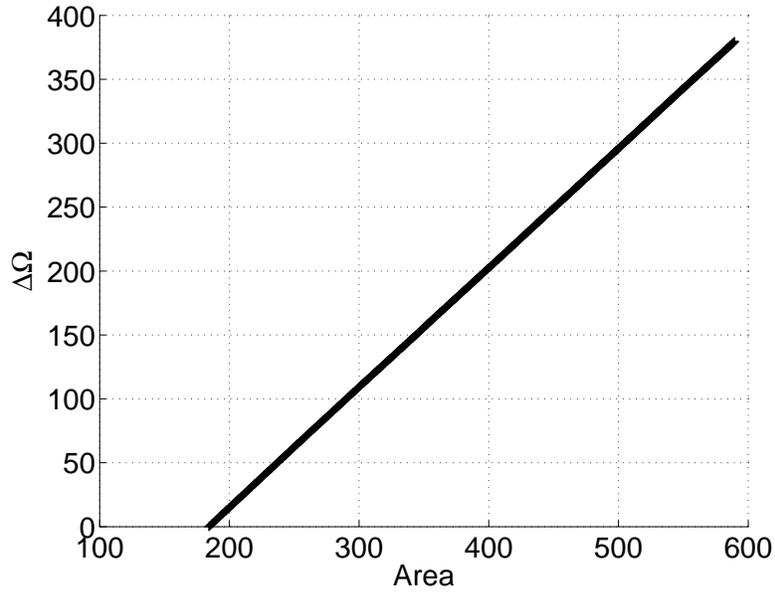}
\end{center}
\caption{Umbrella sampling estimate of the liquid vapor
$\gamma_{LV}$. The measured grand potential is plotted as a
function of the liquid-vapor area. Numerical parameters are
$T/T_c=0.4$ and $\mu/w_{ff}=-4.0$ (corresponding to the
liquid-vapor coexistence). The stiffness $\kappa$ of the biasing
potential is $\kappa=0.01$ and the total system size is $2\times
20^3$ sites.} \label{fig:glvMCumb}
\end{figure}

As shown in figure \ref{fig:glvMCumb}, the measured grand
potential exhibits a linear slope as function of the liquid-vapor
area, as expected. This allows to define unambiguously the
liquid-vapor surface tension. The latter is  "averaged" over the
various directions of the underlying lattice grid. This
"spherical" estimate of the liquid-vapor free energy is compared
in figure \ref{fig:glvMC}
 with the value of the surface free energy along the  [110] direction.

\subsubsection{Contact Angle}
\label{sec:theta}

The contact angle is another necessary ingredient in the classical
description
 of nucleation phenomena on surfaces \cite{barrathansen}.
It is defined  in terms of the various surface free energies,
liquid-vapor (LV), solid-liquid (SL) and solid-vapor (SV),
according to Young's law,
$\cos(\theta)=\frac{\gamma_{SV}-\gamma_{SL}}{\gamma_{LV}}$. This
requires the computation of the solid-liquid and solid-vapor
surface free energies. As for the case of the liquid-vapor surface
free energy, these quantities depend on the specific orientation
of the interface with respect to the underlying lattice.
Technically, planar liquid-solid or vapor-solid interfaces are
constructed and the surface free energies are computed along the
same lines as for the liquid-vapor surface tension (both in the
mean-field and Monte-Carlo calculations).

\begin{figure}
\begin{center}
\includegraphics[height=8cm]{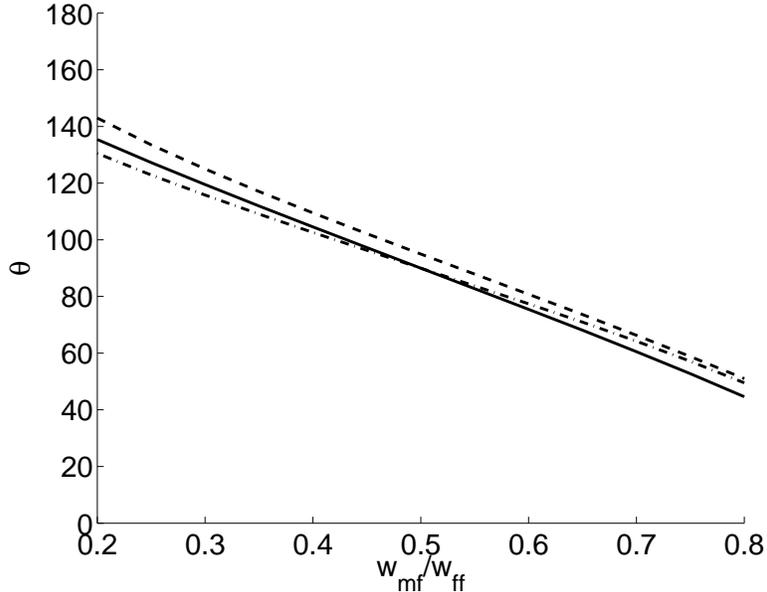}
\end{center}
\caption{Mean field estimate of contact angle along various
directions at $T/T_c=0.4$ as a function of $w_{mf}/w_{ff}$. The
symbols are the same as in   figure \ref{fig:glv}.}
\label{fig:theta}
\end{figure}

The mean field results are displayed in figure \ref{fig:theta}.
The comparison between the contact angle computed for various
directions of the underlying lattice shows only a weak dependence
of the contact angle on the direction of the interface.

\begin{figure}
\begin{center}
\includegraphics[height=4cm]{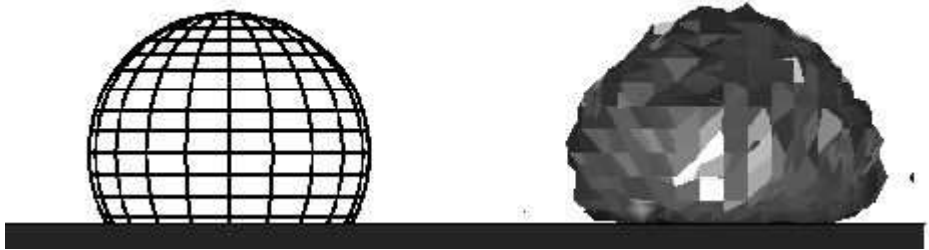}
\end{center}
\caption{Snapshot of a liquid drop growing on a
 planar surface in direction [100] ($w_{mf}/w_{ff}=0.3$, $T/T_c=0.4$, $1500$ liquid sites)
 compared with a macroscopic
  drop with a $120^\circ$ contact angle.
   $\kappa=0.02$. One can check that the underlying BCC
    lattice does not affect the spherical shape of the bubble.}
\label{fig:contactanglebulle}
\end{figure}

\begin{figure}
\begin{center}
\includegraphics[height=8cm]{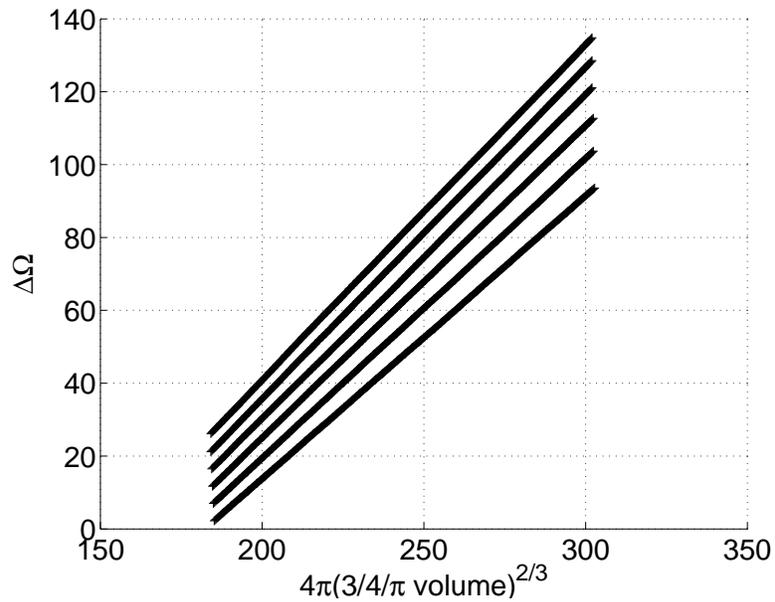}
\end{center}
\caption{Umbrella sampling estimate of the contact angle $\theta$
for a plane normal to direction [100] as a function of the
wettability $w_{mf}/w_{ff}$ at $T/T_c=0.4$. $\glv=0.93515$. From
top to bottom $w_{mf}/w_{ff}=0.20,~0.25,~0.30,~0.35,~0.40,~0.45$.
The measured grand potential of a bubble on a planar surface is
plotted as a function of the power 2/3 of the volume.
$\mu/w_{ff}=-4.0$, $\kappa=0.01$.} \label{fig:thetaMCumb}
\end{figure}

\begin{figure}
\begin{center}
\includegraphics[height=8cm]{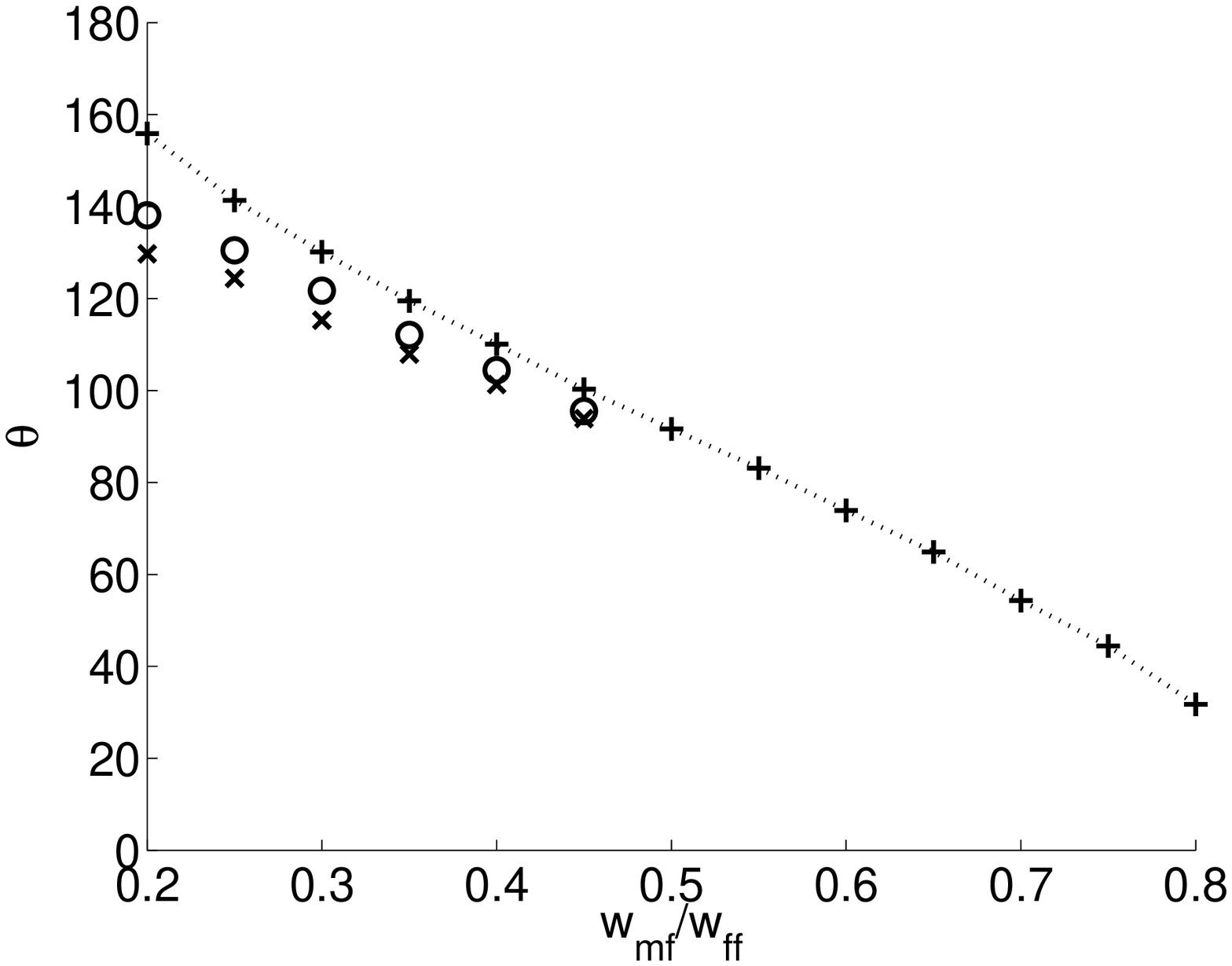}
\end{center}
\caption{Monte Carlo estimate of contact angle along various
directions at $T/T_c=0.4$ as a function of $w_{mf}/w_{ff}$. The
dotted line  corresponds to the
 mechanical
estimate (implemented in the Monte Carlo sampling) in the [110]
direction. Thermodynamic integration estimates are shown for a
[100] (circles) and [110] (crosses) interface.}
\label{fig:thetaMC}
\end{figure}

Beyond mean-field, the previous method based on a biased
Hamiltonian can be used. A liquid bubble is created on a planar
solid surface by adding to the Hamiltonian a penalty associated
with the number of liquid sites (see figure
\ref{fig:contactanglebulle}).
The free energy is computed accordingly from the matching of the
histograms, as in equation (\ref{umb}). The contact angle is
computed by comparing the measured free energy with the prediction
of macroscopic theory. The latter predicts that the excess grand
potential is proportional to   the  $V^{2/3}$, where $V$ is   the
volume of the drop:
 \begin{equation} \Delta \Omega = 4\pi\glv g(\theta)^{1/3}(\frac{3}{4\pi}
V)^{2/3} \end{equation} Knowing $\glv$ from the spherical estimate
in the previous section, the slope of the curve (see figure
\ref{fig:thetaMCumb}) can be used to extract the contact angle,
using $g(\theta)=(2+3\cos(\pi-\theta)-\cos(\pi-\theta)^3)/4$. This
estimate remains however direction dependent, since no averaging
over the orientations of the solid is involved. Two  calculations
for interfaces  in the directions [100] and [110] were performed.
As for the mean-field case, the resulting contact angle is found
to be only weakly dependent on the direction of the interface.
This is shown in figure \ref{fig:thetaMC} where the different
estimates are compared. The small difference between the two
orientations  is likely  due to a change in the solid coordination
number for a liquid site near the interface (4 in [100] direction,
2 in [110] direction).
It is interesting to remark that the contact angles obtained using
the umbrella sampling approach for a droplet are slightly smaller
than those obtained using the mechanical route, which only
involves planar interfaces. This points towards the role of a
nonzero line tension, which has been neglected in the analysis of
the results shown in figure \ref{fig:thetaMCumb}, but will prove
important for the critical nuclei described in the next section,
in which the length of the three phase line is important.

\section{Nucleation Path}
\label{nucleation}

\begin{figure}
\begin{center}
\includegraphics[height=8cm]{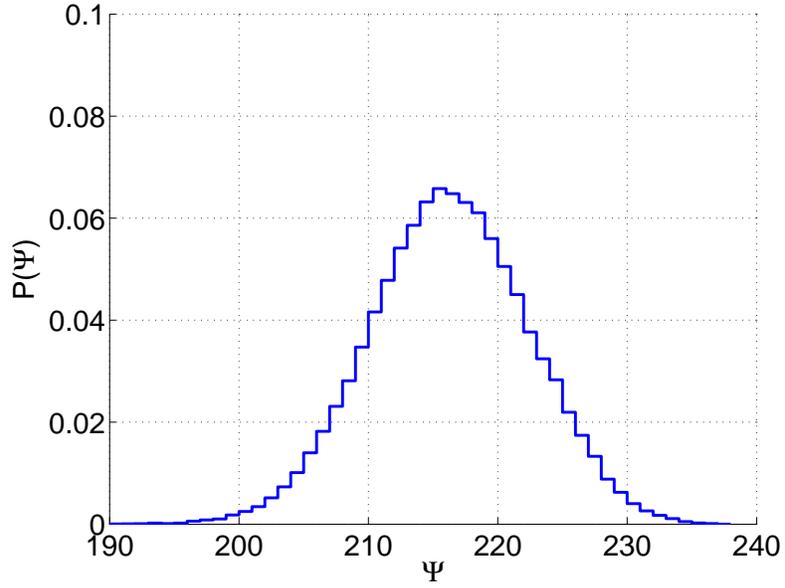}
\end{center}
\caption{State probability distribution as a function of $\Psi$
obtained for a cylindrical pore of radius $5.5$ and length $30$ at
$\mu=-4.0$, $T/T_c=0.4$ and  $w_{mf}/w_{ff}=0.3$. $\Psi_0=230$,
$\kappa=0.005$. The distribution is obtained from 2000 monte carlo
steps, after 1000 steps of averaging. The difference between the
maximum in the probability and the "imposed"  $\Psi_0$ is typical
of an unstable situation.} \label{fig:distribproba}
\end{figure}

\begin{figure}
\begin{center}
\includegraphics[height=8cm]{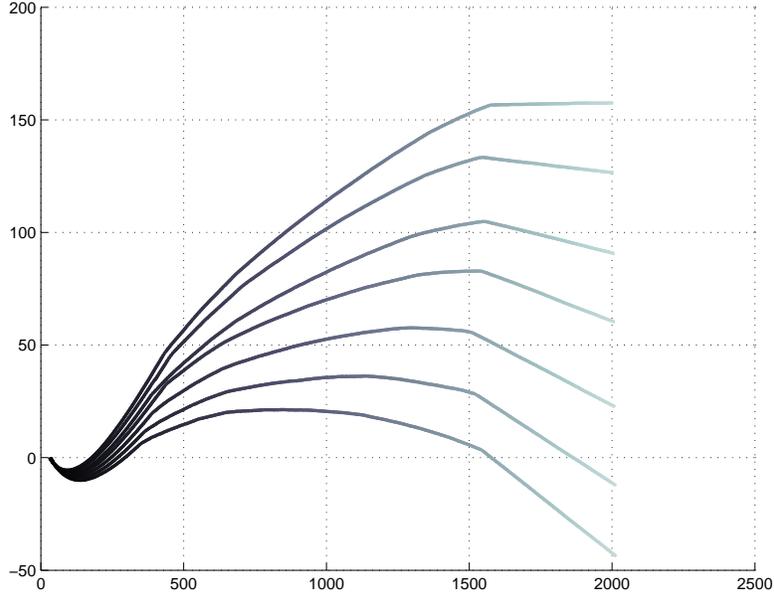}
\end{center}
\caption{Grand potential as a function of $\Psi$ for a cylindrical
pore of radius $5.5$ and length $30$. From top to
bottom,$\mu=-3.88$, $\mu=-3.90$, $\mu=-3.92$, $\mu=-3.94$,
$\mu=-3.96$, $\mu=-3.98$, and $\mu=-4.0$. $\kappa=0.01$. The
sampling windows for $\Psi$ are shifted by   10 units. The final
state corresponds to a partially filled pore containing two
liquid-vapor interfaces. } \label{fig:umbrellaomega}
\end{figure}

We now focus on the nucleation path for  capillary desorption. To
this end, we use the previously described umbrella sampling
technique for a "biased" system \cite{frenkel}. An order parameter
$\Psi$ is defined as the number of vapor sites (with density less
than $0.5$) and the Hamiltonian is biased by adding a term in
$\frac{1}{2}\kappa(\Psi-\Psi_0)^2$ where $\Psi_0$ is the target
value for sampling. Typical values of $\kappa$ are $0.005$. State
probability distributions $P(\Psi)$ are obtained over a window of
range $\Delta \Psi=30$ (see figure \ref{fig:distribproba}).
The nucleation path is sampled by starting with a filled pore
($\Psi_0=0$) and progressively increasing the number of empty
sites. For each value of the order parameter, $1000$ Monte Carlo
steps where used for equilibration and $2000$ steps for
statistics. The final configuration serves as initial
configuration for the next order parameter value. Grand potential
curves are estimated from the order parameter histogram, using
equation (\ref{umb}).
As usual, a matching procedure between different order parameter
windows is used to obtain the free energy curve. The results are
shown in figure \ref{fig:umbrellaomega} for different values of
the chemical potential for a cylindrical pore of radius $5.5$ and
length $30$.

A metastability limit of the  liquid filled pore is reached
  around $\Psi=1550$.
Once the vapor bubble fills the whole cylinder radius, it adopts a
"cylinder like" shape, whose length increases linearly with
$\Psi$. The nucleation barrier is defined as the difference
between maximum grand potential and its value in the liquid
metastable state. $\Psi=100$ (depending on the chemical
potentiel), and corresponds to the presence of a local vapor
region near the hydrophobic wall. Such regions are subcritical
vapor bubbles that can lead to nucleation. Increasing the  order
parameter $\Psi_0$, a vapor bubble grows in contact with the wall
 and suddenly turns  into a vapor cylinder terminated by two
 spherical caps.

\begin{figure}
\begin{center}
 \includegraphics[width=6cm]{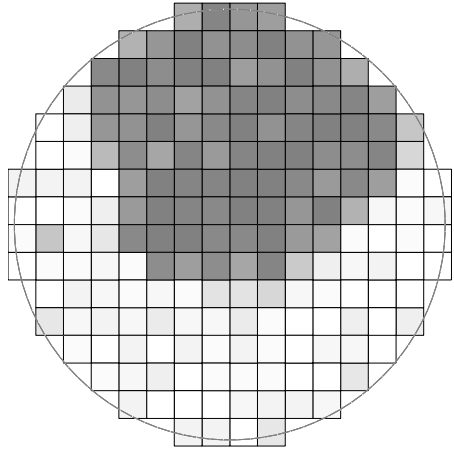}
\end{center}
\caption{Cross section of a typical umbrella sampling snapshot,
before the energy barrier is crossed, for a cylindrical pore of
radius $5.5$ and length $30$. The cross section is taken in the
plane of symmetry of the bubble. An iso-density surface is
represented for $\mu=-3.94$, $\Psi_0=1535$.}
\label{fig:snapshotsa}
\end{figure}

\begin{figure}
\begin{center}
\includegraphics[width=10cm]{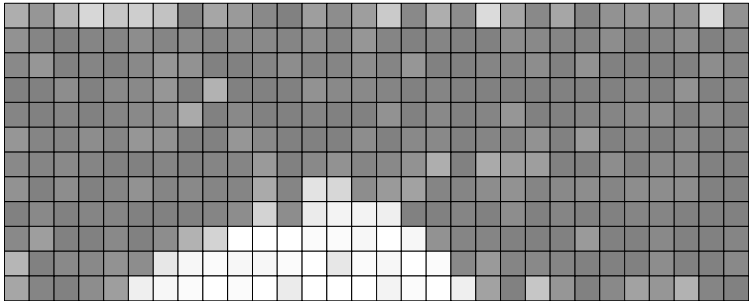}
\end{center}
\caption{Same as in figure \ref{fig:snapshotsa}, but perpendicular
to the axis of the cylinder. Again the cross section is taken in
the plane of symmetry of the bubble. Note that both views are
representations of the same nucleus, which makes evident the
saddle shape of the critical nucleus.} \label{fig:snapshotsb}
\end{figure}

This approach also yields the shape of the critical nucleus.
 A typical example is shown on figures \ref{fig:snapshotsa},\ref{fig:snapshotsb}.
This shows that the nucleation process occurs via the creation of
a vapor bubble on the wall. It is important to emphasize that the
nucleus breaks the cylindrical symmetry of the system, in contrast
to simple expectations \cite{lefevrecnt}.

\section{Nucleation Barrier}

We now gather the results for the energy barrier measured in the
simulations, as a function of the various thermodynamic
parameters. The temperature of the system is $T/T_c=0.4$ giving an
averaged liquid-vapor surface tension $\glv=0.93515$ from the
thermodynamic integration estimate. The wettability of the
confining pore is $w_{mf}/w_{ff}=0.3$. The  contact angle is
estimated to be   $121.7^\circ$ in [100] direction and
$115.3^\circ$ for the [110] interface (see section
\ref{sec:gammaLV}).

\begin{figure}
\begin{center}
\includegraphics[height=8cm]{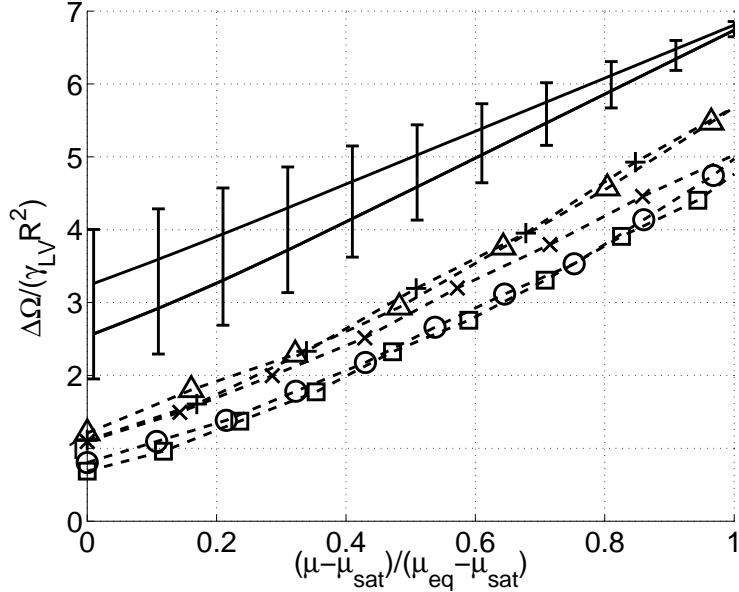}
\end{center}
\caption{Reduced energy barrier for cylindrical pores of radius
$R=3.5$ (circle), $R=4$ (square), $R=4.5$ (cross), $R=5$
(triangle), $R=5.5$ (plus) compared to macroscopic results for a
vapor bubble using a $120^\circ$ (lower full line) and $115^\circ$
(upper full line) contact angle. The error bars corresponds to the
same macroscopic estimates,  with  contact angle between
$110^\circ$ and $125^\circ$. Dotted lines
 are guides to the eye.
}
\label{fig:umbrellabarred}
\end{figure}

We plot in figure \ref{fig:umbrellabarred} the results for the
reduced nucleation barrier $\Delta\Omega/\gamma_{LV} R^2$,  $R$
being  the pore radius. The nucleation barrier is plotted against
the "metastability ratio" $(\mu-\mu_{sat})/(\mu_{eq}-\mu_{sat})$,
where $\mu$ is the actual chemical potential, $\mu_{sat}$ is the
bulk coexistence chemical potential, and $\mu_{eq}$ is the
equilibrium chemical potential for the capillary evaporation in
the considered pore of radius $R$. The latter is computed
independently, by preparing a configuration in which vapor and
liquid phases, separated by a meniscus, are coexisting inside the
pore.

From a macroscopic point of view, the excess grand potential
between a pore filled of liquid and a pore containing a vapor
nucleus can be expressed as
\begin{equation}
\Delta\Omega=V_{V}(P_L-P_V) +\gamma_{LV} A_{LV} + (\gamma_{SV}-\gamma_{SL})A_{SV}
\end{equation}
Here $V$ is the volume of vapor phase and $A_{SL}$, $A_{SV}$,
$A_{LV}$ are the solid-liquid, solid-vapor and liquid-vapor
surface areas. Using reduced quantities
$\tilde{V}_V=\frac{V_V}{R^3}$,
$\tilde{A}_{LV}=\frac{A_{LV}}{R^2}$,
$\tilde{A}_{SV}=\frac{A_{SV}}{R^2}$, the definition of the contact
angle $\cos(\theta)=\frac{\gamma_{SV}-\gamma_{SL}}{\gamma_{LV}}$,
and introducing  Kelvin's radius $R_K=\frac{\glv}{P_L-P_V}$, one
obtains
\begin{equation}
\frac{\Delta\Omega}{\glv R^2}~=~\frac{R}{R_K}~\tilde{V}~+~\tilde{A}_{LV}~-~\cos(\pi-\theta)~\tilde{A}_{SV}
\label{eqn:barredmacro}
\end{equation}
The parameter $R/R_K$ can be related  to the metastability ratio
using
\begin{equation}
\frac{\mu-\mu_{sat}}{\mu_{eq}-\mu_{sat}}=\frac{P_L-P_{sat}}{P_L^{eq}-P_{sat}}\simeq\frac{R}{2
R_K \vert \cos(\theta)\vert}
\end{equation}
On the other hand, the reduced areas and volumes, $\tilde{A}_{LV}$, $\tilde{A}_{SV}$ and $\tilde{V}_V$,
do depend on the specific geometry and morphology of the critical nucleus. In  a previous paper,
ref. \cite{lefevrecnt}, we have proposed a detailed calculation of these quantities and obtained
the corresponding energy barrier for a nucleus with an asymetric shape, as observed in the present simulations
(see figures \ref{fig:snapshotsa} and \ref{fig:snapshotsb}). We refer to this paper for further details
on these calculations.
We only quote a simple and convenient approximation for the energy barrier, which writes
\be
\Delta\Omega~=~(P_L-P_V) K_1 R^3+\glv K_2 R^2
\label{approx} \ee with the constants $K_1=4.18$ and  $K_2=2.12$
for $\theta=120^\circ$ \cite{lefevrecnt}.

Figure \ref{fig:umbrellabarred} shows a comparison between the
macroscopic estimate of the nucleation barrier, computed from
classical capillarity as described  in
\cite{lefevrecnt}, with results obtained from Monte Carlo
simulations for cylindrical pores of radius $R=3.5$, $R=4$,
$R=4.5$, $R=5$ and $R=5.5$. The classical capillarity estimate for
the reduced barrier does not depend on pore size, as is obvious
from dimensional arguments.
Even allowing for some flexibility in the value of the contact
angle, it is clear that that  the macroscopic approach
overestimates the nucleation barrier,  and that the reduced energy
barrier obtained from simulation depends on the  size of the pore.
The good agreement in the slopes
 suggests however that  this discrepancy can be corrected by
 a simple shift of  the classical capillarity estimate of the activation
 energy, which should depend on  pore size and be
independent of the chemical potential. As already proposed by
Lefevre et al. \cite{lefevrecnt}, such a shift can be obtained by
invoking the role of line tension,  which adds to the nucleation
barrier a term proportional to the length of the three phase line
in the critical nucleus. Such a  line term writes
$\tau^{*}\tilde{L}_{SLV}/(\glv R)$, where $\tilde{L}_{SLV}$ is the
length of the three phase line.

\begin{figure}
\begin{center}
\includegraphics[height=8cm]{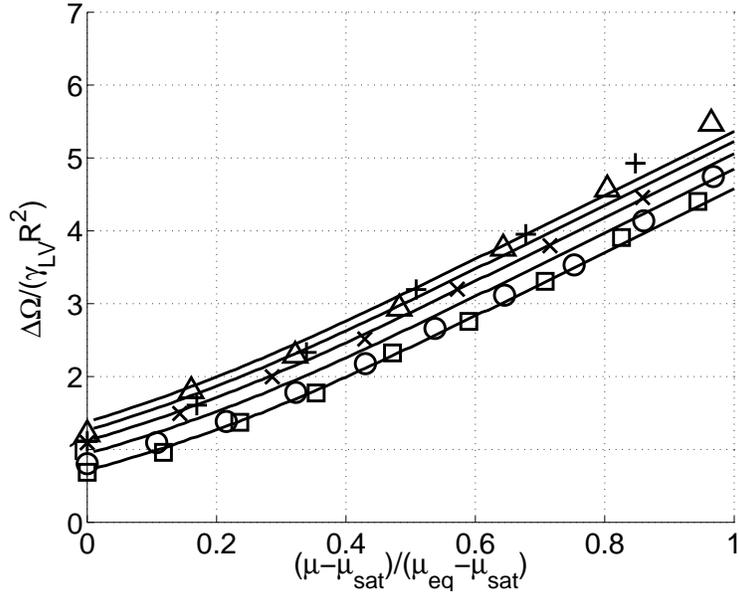}
\end{center}
\caption{Macroscopic reduced energy barriers for cylindrical pores
using a $120^\circ$ contact angle and a  line tension $\tau^*=
-0.55 w_{ff}/b$
 (full lines) compared with Monte Carlo simulations. From bottom
 to top $R=3.5$, $R=4$, $R=4.5$, $R=5$ and $R=5.5$. The symbols are the same as in figure
\ref{fig:umbrellabarred}.} \label{fig:umbrellabarredlambda}
\end{figure}

Such a term is easily taken into account in the previous macroscopic
description, as described in ref. \cite{lefevrecnt}.  The
results, displayed in figure \ref{fig:umbrellabarredlambda}, show
that a value of the line tension $\tau^{*}/w_{ff}=-0.55/b$ ($b$
being the unit length of the lattice) allows to obtain a very good
agreement between the measured free energy barriers and the
macroscopic estimate. Again we quote the corresponding convenient
approximation of the free energy barrier with the line tension term
included (see Eq. (\ref{approx})):
\be
\Delta\Omega~=~(P_L-P_V) K_1 R^3+\glv K_2 R^2+\lambda K_3 R
\label{approx2}
\ee
with $K_3=12.43$ for $\theta=120^\circ$ \cite{lefevrecnt}.

A few comments are in order. First the negative sign of the line
tension is in agreement with previous observations
\cite{talanquer,lefevrecnt}. Moreover, the order of magnitude is
consistent with these previous estimates. Indeed, using  a typical
microscopic length $b\simeq 1$ nm and the value of the critical
temperature of water, $T_c \simeq 675^\circ$ K, we obtain
$\tau~\sim 10^{-11}~J/m$ in good agreement with experimental
datas.

\section{Conclusion}
In this paper a careful estimate of the nucleation barriers for
capillary evaporation inside an hydrophobic pore has been
proposed.  Monte Carlo simulations have shown that capillary
evaporation occurs via the nucleation of a vapor bubble at the
wall of the cylinder pore. Therefore the critical nucleus does not
exhibit the cylindrical symmetry of the cylinder. We have shown
moreover that a macroscopic estimate of the free energy is
consistent with the measured free energy barriers in a hydrophobic
cylindrical pore, provided a contribution from the  line tension
is included. This conclusion is consistent previous studies
\cite{talanquer,lefevrecnt}, and should motivate more direct
determination of line tensions in such systems, using experimental
or numerical tools.


\appendix
\section{Mechanical calculation of interfacial tension}
\label{ann:tension} Consider a $3D$ system of volume $V$ in
contact with an external reservoir of temperature T and chemical
potential $\mu$ and described  by the Hamiltonian ${\mathcal H}_0$
and partition function $\Theta_0$ The system is assumed to contain
 a liquid-vapor  interface of area $A$ normal to the z axis. By transformation
$x\to (1+\lambda)x$, $y\to (1+\lambda)y$ and $z\to (1-2\lambda)z$,
$dV={\mathcal O}(\lambda^2)$ and $dA=2\lambda A$. the Hamiltonian
becomes, to first order in $\lambda$,
\begin{equation}
{\mathcal H}={\mathcal H}_0+2\lambda{\mathcal H}_1
\end{equation}

From linear response, the new  partition function is given by
\begin{equation}
\Theta=\Theta_0+\delta\Theta= \int \exp{-\beta({\mathcal
H}_0+2\lambda{\mathcal H}_1)}\,d\Gamma \approx \int
\exp{-\beta{\mathcal H}_0}(1-2\beta\lambda{\mathcal H}_1)
\,d\Gamma
\end{equation}
Thus
\begin{equation}
\delta\Theta=-2\beta\lambda<{\mathcal H}_1>\Theta_0
\end{equation}
The change $d\Omega$ in the grand potential  can be calculated
using the  macroscopic definition $d\Omega=\gamma
dA=2\lambda\gamma A$ or the  microscopic approach
$\Omega=-k_BTlog\Theta~\Rightarrow~d\Omega=-k_BT\delta\Theta/\Theta
=2\lambda<\mathcal{H}_1>$ leading to the mechanical expression of the
interfacial tension :
\begin{equation}
\gamma A=<{\mathcal H}_1>
\end{equation}

Dealing with the Hamiltonian introduced in section
\ref{sect:model} on a BCC lattice with cubic  axes
$\vec{x},\vec{y},\vec{z}$, we write the rescaling of a system with
an interface in the plane $(\vec{u}=(1,1,0),\vec{z})$ normal to
$\vec{v}=(-1,1,0)$ as $\vec{u}\to (1+\lambda)\vec{u}$, $\vec{z}\to
 (1+\lambda)\vec{z}$ and$\vec{v}\to (1-2\lambda)\vec{v}$. This is
taken into account in the  Hamiltonian by assuming a dependence of
$w_{ff}$ and $w_{mf}$ as the inverse square length between nearest
neighbors sites \cite{restagno,rowlinsom}. Thus ${\mathcal H}_1$
is obtained as :
\begin{eqnarray}
&{\mathcal H}_1&=\sum_{<ij>_{(\vec{u},\vec{z})}}
 w_{ff}\rho_i\rho_j+w_{mf}[\rho_i(1-\eta_j) +(1-\eta_i)\rho_j]\nonumber\\
& & -\sum_{<ij>_{(\vec{v},\vec{z})}}w_{ff}\rho_i\rho_j+w_{mf}[\rho_i(1-\eta_j) +(1-\eta_i)\rho_j]
\end{eqnarray}
where $<ij>_{(\vec{u},\vec{z})}$ (resp $<ij>_{(\vec{v},\vec{z})}$)
 are interaction in the $(\vec{u},\vec{z})$ (resp $(\vec{v},\vec{z})$) plane.

\bibliographystyle{unsrt}

\begin{thebibliography}{10}

\bibitem{maibaum}
Maibaum L., Chandler D., J. Phys. Chem. B, \textbf{107},
1189~(2003).

\bibitem{hummer}
Hummer G., Rasaiah J.C., Noworyta J.P. , Nature, \textbf{414},
188~(2001).

\bibitem{bolhuis}
Bolhuis P.G., Chandler D. , J. Chem. Phys., \textbf{113},
8154~(2000).

\bibitem{barrat}
Barrat J-L.,  Bocquet L., Phys. Rev. Lett., \textbf{82},
4671~(1999).

\bibitem{eros01}
Eroshenko V.A.~,Regis R.C., Soulard~M., Patarin J., J. Am. Chem.
Soc., \textbf{123}, 8129~(2001).


\bibitem{lefevre}
Martin T., Lefevre B. ~\emph{et al}, Chem. Commun., \textbf{1},
24~(2002).

\bibitem{gelb}
Gelb L.D., Gubbins K.E., Radhakrishnan R., Sliwinska-Bartkowiak
M., Rep. Prog. Phys., \textbf{62}, 1573~(1999).


\bibitem{chandlernature}
 Chandler, D. Nature, \textbf{417}, 491 (2002)


\bibitem{maibaum2}
Maibaum, L., A. R. Dinner and D. Chandler. J. Phys. Chem. B (In
Press)(2004).


\bibitem{tenwolde}
TenWolde  P. R, D. Chandler D., Proc. Nat. Ac. Sci.,\textbf{99},
6539-6543 (2002).


\bibitem{huang2}
Huang, D.M,  Chandler D., J. Phys. Chem. B, \textbf{106},
2047-2053 (2002).


\bibitem{huang}
Huang, D. M. and D. Chandler, Phys. Rev. \textbf{E 61}, 1501-1506
(2000).


\bibitem{kierlikjpcm}
Kierlik E. , Monson P.A., Rosinberg M.L., Tarjus G., J. Phys.:
Condens. Matter, \textbf{14}, 9295 (2002).



\bibitem{kierlikprl}
Rosinberg M.L., Kierlik E., Tarjus G., Europhys. Lett. {\bf 62},
377 (2003)


\bibitem{lefevrecnt}
Lefevre B., Saugey A., Barrat J-L. Bocquet L., Charlaix E., Vigier
G., Gobin. P-F., J. Chem. Phys. {\bf 120}, 4927-4938 (2004)



\bibitem{lum}
Lum ~K., Chandler D., Int. J. Thermo.,\textbf{19}, 845-855 (1998).



\bibitem{lum2}
K.~Lum~A.~Luzar, Phys. Rev. E, 56, 6283~(1997).



\bibitem{restagno}
Restagno F., Bocquet L., Biben T., Phys. Rev. Lett., \textbf{84},
2433~(2000).

\bibitem{talanquer}
Talanquer V., Oxtoby D.W., J. Chem. Phys., \textbf{114},
2793~(2001).


\bibitem{bolhuis2}
Bolhuis, P. G.,  Chandler D., Dellago C., Geissler P., Ann. Rev.
of Phys. Chem.,\textbf{59}, 291-318 (2002).

\bibitem{dellago}
Dellago, C. and D. Chandler, Lect. Notes in Phys., \textbf{605},
321-333(2002).

\bibitem{kornev}
Kornev K.G., Neimark A.V., Adv. Colloid Interface Sci.,
\textbf{96}, 143~(2002).

\bibitem{kierlikpccp}
Kierlik E., Rosinberg M.L., Tarjus G., Viot P., Phys. Chem. Chem.
Phys., \textbf{3}, 1201~(2001).


\bibitem{sarkisov}
Sarkisov L., Monson P.A., Phys. Rev. E, \textbf{65},
011202~(2001).




\bibitem{chandler}
D.~Chandler, "Introduction to Modern Statistical Mechanics",
Oxford University Press, Oxford 1987.



\bibitem{frenkel}
Frenkel D., Smit B., "Understanding Molecular Simulation",
Academic Press, New-York 2001.

\bibitem{rowlinsom}
Rowlinson J.S., Widom B., "Molecular Theory of Capillarity" Oxford
University Press,  Oxford, 1989.


\bibitem{barrathansen}
Barrat J-L., Hansen J-P. "Basic concepts for simple and complex
fluids", Cambridge University Press, Cambridge 1983.


\end{thebibliography}

\end{document}